\documentclass[10pt]{article}
\usepackage{amsmath,amsfonts,amssymb,amscd,amsxtra,amsthm}

\usepackage{graphicx}
\usepackage{epsfig}
\usepackage{bm}
\vspace{-8ex}
  \date{}
\begin{document}
\title{Parameterisation space for Cornell potential in a QCD potential model}
\author{ $^{1}$Krishna Kingkar Pathak, $^{2}$ Satyadeep Bhattacharya and $^{3}$ Tapashi Das \\
$^{1}$Department of Physics,Arya Vidyapeeth College,Guwahati-781016,India\\
$^{2}$ Impact Hub Institute for Competitive Examination, Dimapur- 797112, Nagaland, India\\
$^{3}$Department of Physics, Madhab Choudhury College, Barpeta- 781301, India\\
e-mail:tapashimcc@gmail.com}
\maketitle
\begin{abstract}
We make a critical analysis on the free parameters of the Cornell  potential $-\frac{4\alpha_{s}}{3r}+br+c$  and provide a parameterisation space for the strong coupling constant $\alpha_{s}$ and the constant shift `c' for choosing linear part as perturbation in the potential model. In the analysis of  heavy-light mesons $(D,D_{s},B,B_{s}~ \text{and} ~B_{c})$, we have found  a wide range of values for the coupling constant i.e $0.20 \leq \alpha_{s}\leq 0.64$ with $-1.2 \leq c \leq -0.66$ which can be used to treat the confining part as perturbation.
 \end{abstract} 
\section{Introduction}
 QCD potential between a quark and anti quark has been  the first important ingredient of phenomenological model to study Hadron Physics. The potential model is found to be successful in providing both qualitative and quantitative description of hadron spectrum and it's decay modes. To deal with a potential model, the choice of the correct QCD potential is the most important ingredient for its success. There are several acceptable potentials in QCD, which is based upon the two important facts of QCD i.e. the confinement and asymptotic freedom of quarks. To study quark-anti quark bound states, some of the accepted and commonly used potentials are, Cornell potential\cite{cornell}: $-\frac{A}{r}+Br+c$, Power law potential\cite{sameer,tsong}: $-Ar^\alpha+Dr^\beta+c$, Logarithmic potential\cite{log}: $A+B~lnr$, Richardson potential\cite{richardson}:$ Ar-\frac{B}{rln\frac{1}{\lambda^r}}$ etc.\\
 
However, the potential parameters in the different potentials as well as in the different models are found to vary within a noticeable range. For example, the value of `c' is found to vary from model to model. In the work of relativistic quark model, Faustov et. al. has used $c=-0.3 GeV$ \cite{faustov}, E. Eichten et al. in ref. \cite{99} has taken it to be $c=0.50805 GeV$, Mao Zhi Yang has taken $c=-0.19GeV$ \cite{mao}, Scora and Isgur considered $c=-0.81 GeV$ in ref. \cite{scora}, whereas Grant and Rosner \cite{grant} considered a large negative value for $c=-1.305GeV$ in a power law potential. The value of the strong running coupling constant $\alpha_{s}$ appearing in the  QCD potential has also a wide range from $0.22$ to $0.64$ in different theoretical works as well as in the different schemes like $\overline{MS}$, $V$-scheme etc. \cite{groom, lahkar, godfrey, richard}. \\

In this work, we review our previous work \cite{nsbijmpa, pramana} put forward some comments on the perturbation theory of choosing the linear part of the Cornell potential as perturbation and provide a parameterisation space for the strong coupling constant $\alpha_{s}$ and the constant shift $c$. While taking the Coulomb part as parent and linear as perturbation one important point is to be noted that for this choice, the perturbation is possible only for very small value of `b'. In infinite mass limit, the value of `b' is $\le$ 0.03 $GeV^2$ \cite{nsbijmpa} which are much smaller than the value of `b' in charmonium spectroscopy ($\approx 0.183 GeV^2$).\\

The paper is organized  as follows, in the section $2$, we describe the Cornell potential and its parameters. In Section $3$, we describe the QCD potential model with linear part as perturbation. In section $4$, we discuss the constraints on the strong coupling constant $\alpha_{s}$ and $c$. Section $5$ contains the conclusions.

 \section{The Cornell Potential and Perturbation}
 For mesons, the one gluon exchange contribution between quark and antiquark is given by the coulombic potential,
\begin{equation}
V(r)=-\frac{4\alpha_s}{3r}.
\end{equation}

Here, $-\frac{4}{3}$ is due to the SU(3) color factor and $\alpha_s$ is the strong coupling constant.\\

For large distance, we have  confinement of quarks. The confining potential is given by 
\begin{equation}
V(r)=br,
\end{equation}

where $b$ is known as confinement parameter and phenomenologically $b=0.183 GeV^2$ \cite{pramana}.\\

Consdering the two effects,  Cornell potential is written as:
\begin{equation}
V(r)=-\frac{4\alpha_s}{3r}+br+c
\end{equation}

which is the sum of coulombic and linear potential with a scale factor `c'. Both the potentials play decisive role in the quark dynamics and their separation is not possible. Besides there is no appropriate small parameter so that one of the potential within a perturbation theory can be made perturbative.\\
 
With the Cornell Potential, one cannot solve the Schrodinger equation in quantum mechanics except for some simple models. Therefore, physicists opt for developing efficient approximate methods. Perturbation theory is one of the helpful tools to get an approximate wavefunction with the Cornell potential. In fact, perturbation theory is considered to be one of the approximate methods which most appeals to intuition.\\ 

However,in perturbation theory, one has to check the convergence of the series which appears in the procedure. If the rate of convergence of the perturbation series is sufficiently high then we may expect accurate results.\\

The advantage of taking Cornell potential for study is that it leads naturally two choices of ``parent" Hamiltonian, one based on the Coulomb part and the other on the linear term, which can be usefully compared. It is expected that, in choosing the perturbative part of the potential a dominant role is played by critical $ r_{0} $ where the potential $ V(r)=0 $. Aitchison and Dudek in ref.\cite{AITCHISION} put an argument that if the size of a state measured by $\langle r\rangle < r_{0}$, then the Coulomb part as the ``Parent" will perform better and if not so the linear part as ``parent" will perform better. The Aitchison's work also showed the results that with Coulombic part as perturbation(VIPT), bottomonium spectra are well explained than Charmonium where as Charmonium states are well explained with linear part as parent. It becomes noteworthy in this context that the critical distance $ r_{0} $ is not a constant and can be enhanced by reducing $b$ and $c$ or by increasing $\alpha_{s}$.

\section{The QCD Potential Model}
In this work, a specific potential model with linear part as perturbation is taken into consideration. For completeness and proper reference we put the last modified version of the wavefunction with coulombic part as parent as in ref.s \cite{pramana,NSB, CPL}.\\

The non-relativistic predictions of potential models with a non-relativistic Hamiltonian for the heavy-light and heavy-heavy mesons are found to be in fair agreements with the updated theoretical, experimental and lattice results. Hence, we start with the ground state ($l=0$) spin independent non-relativistic Fermi-Breit Hamiltonian (without the contact term),

\begin{equation}
H=-\frac{\nabla^{2}}{2\mu}-\frac{4\alpha_s}{3r}+br+c. 
\end{equation}

With the linear term `$br+c$' as perturbation and using Dalgarno method, the wave function in the model is obtained as \cite{ pramana, NSB, CPL,NSB2009}

\begin{equation}
\label{wf}
\psi_{rel+conf}\left(r\right)=\frac{N^{\prime}}{\sqrt{\pi a_{0}^{3}}} e^{\frac{-r}{a_{0}}}\left( C^{\prime}-\frac{\mu b a_{0} r^{2}}{2}\right)\left(\frac{r}{a_{0}}\right)^{-\epsilon},
\end{equation}

where

\begin{equation}
N^{\prime}=\frac{2^{\frac{1}{2}}}{\sqrt{\left(2^{2\epsilon} \Gamma\left(3-2\epsilon\right) C^{\prime 2}-\frac{1}{4}\mu b a_{0}^{3}\Gamma\left(5-2\epsilon\right)C^{\prime}+\frac{1}{64}\mu^{2} b^{2} a_{0}^{6}\Gamma\left(7-2\epsilon\right)\right)}},
\end{equation}

\begin{equation}
C^{\prime}=1+cA_{0}\sqrt{\pi a_{0}^{3}},
\end{equation}

\begin{equation}
\mu=\frac{m_{i}m_{j}}{m_{i}+m_{j}},
\end{equation}

\begin{equation}
a_{0}=\left(\frac{4}{3}\mu \alpha_{s}\right)^{-1},
\end{equation}

\begin{equation}
\epsilon=1-\sqrt{1-\left(\frac{4}{3}\alpha_{s}\right)^{2}}.
\end{equation}

The QCD potential is taken as
\begin{equation}
V\left(r\right)=-\frac{4}{3r}\alpha_{s}+br+c.
\end{equation}

Here $A_{0}$ is the undetermined factor appearing in the series solution of the Schr\"odinger equation.                                                                                                                                                                  The term $\left(\frac{r}{a_{0}}\right)^{-\epsilon}$ in  equation (\ref{wf}) is the Dirac factor and was introduced to incorporate relativistic effect \cite{CPL,JJ,QFT}. It is to be noted that the factor $\left(\frac{r}{a_{0}}\right)^{-\epsilon}$ to the wave function (\ref{wf}) develops a singularity at $r \to 0$ and the masses of scalar mesons become negative, therefore the term can be omitted from the wave function as dicussed in ref. \cite{lahkar}.\\

Similarly, the wave function with coulombic part `$-\frac{4\alpha_s}{3r}+c$' as perturbation and linear part as parent (upto O(4)) is given by \cite{tdijp},

\begin{equation}
\psi_{rel+lin}(r)=\frac{N^{\prime\prime}}{r} \left[1+A_1(r)r+A_2(r)r^2+A_3(r)r^3+A_4(r)r^4\right]A_i[\rho_1 r+\rho_0] \left( \frac{r}{a_0}\right) ^{-\epsilon}
\end{equation}

where $A_i[r]$ is the Airy function \cite{23} and $N^{\prime\prime}$ is the normalization constant
\begin{equation}
N^{\prime\prime}= \frac{1}{\left[ \int_0^{\infty} 4 \pi \left[1+A_1(r)r+A_2(r)r^2+A_3(r)r^3+A_4(r)r^4\right]^2 \left( A_i[\rho_1 r+\rho_0]\right) ^2 \left( \frac{r}{a_0}\right) ^{-2\epsilon}dr\right]^{\frac{1}{2}} }.
\end{equation}

The co-efficients of the series solution as occured in Dalgarno's method of perturbation, are the functions of $\alpha_s, \mu, b$ and c:

\begin{equation}
A_1=\frac{-2\mu \frac{4\alpha_s}{3}}{2\rho_1 k_1+\rho_1^2 k_2}
\end{equation}

\begin{equation}
A_2=\frac{-2\mu (W^1-c)}{2+4 \rho_1 k_1+ \rho_1^2 k_2}
\end{equation}

\begin{equation}
A_3=\frac{-2\mu W^0 A_1}{6+6 \rho_1 k_1+ \rho_1^2 k_2}
\end{equation}

\begin{equation}
A_4=\frac{-2\mu W^0 A_2+2\mu b A_1}{12+8 \rho_1 k_1+ \rho_1^2 k_2}
\end{equation}

The different parameters are given by,
\begin{equation}
\rho_1=(2\mu b)^{\frac{1}{3}}
\end{equation}

\begin{equation}
\rho_0=-\left[ \frac{3\pi (4n-1)}{8}\right] ^{\frac{2}{3}}
\end{equation}

(In our case n=1 for ground state)

\begin{equation}
k_1=1+\frac{k}{r}
\end{equation}

\begin{equation}
k=\frac{0.3550281-(0.2588194) \rho_0}{(0.2588194) \rho_1}
\end{equation}

\begin{equation}
k_2=\frac{k^2}{r^2}
\end{equation}

\begin{equation}
W^1=\int \psi^{(0)\star} H^{\prime} \psi^{(0)} d\tau
\end{equation}

\begin{equation}
W^0=\int \psi^{(0)\star} H_0 \psi^{(0)} d\tau.
\end{equation}

Let us now discuss the constraints on the parameters $\alpha_{s}$ and $c$ in the model.

\section{Constraints from two points of view}
To evaluate a narrow range of the free parameter in the model, we consider the two constraints; one from the expectation value and the other from the convergence point of view in the model.

\subsection{From the condition of expectation value}

Considering the argument of Aitchison and Dudek \cite{AITCHISION} $\langle r\rangle < r_{0}$ to treat the linear part as perturbation, 
we should get the size of a state measured by $\langle r\rangle < r_{0}$, where $r_{0}$ is the critical distance at which $V(r_{0})=0$. Now,
\begin{equation}
\nonumber
\langle r\rangle_{coul} = \int  \psi^{*} r \psi dr =\frac {3a_{0}} {2}=r_{1}(say)
\end{equation}
and the critical distance $r_{0}$ at which $V(r_{0})=0$ can be obtained by the relation \\
\begin{equation}
\nonumber
br_{0}^{2} +cr_{0}-\frac{4\alpha_{s}}{3}=0.
\end{equation}
The variation of $r_{1}$ and $r_{0}$ with the model parameters can be easily studied from the above relations and the results are tabulated in Table 1. From the results it is clear that to treat linear part as perturbation, the value of upper scale of $\alpha_{s}$ is different for different mesons. The maximum value for $B$ meson is found to be $\alpha_{s}=0.64$ beyond which the condition $\langle r\rangle < r_{0}$ invalid.
%\begin{tabular}{|c|c|c|c|c|c|c|c|}
%\hline Mesons &\multicolumn{2}{c|}{ \alpha_{s}=0.65 } &\multicolumn{2}{c|}{\alpha_{s}=0.5}  &\multicolumn{2}{c|}{\alpha_{s}=0.4 }  \\  
%\hline  &  &  &  \\ 
%\hline  &  &  &  \\ 
%\hline  &  &  &  \\ 
%\hline  &  &  &  \\ 
%\hline  &  &  &  \\ 
%\hline 
%\end{tabular} \\

\begin{table}[h]
\begin{center}
\caption{ Values of $r_{1}$ and $ r_{0}$ for different mesons with $m_{u/d}=0.33 $ $GeV,$ $ m_{s}=0.483 $ $GeV,$ $m_{c}=1.55 GeV$,$ m_{b}=4.97 GeV,$ $b=0.183 GeV^{2}$ and $cA_{0}=1 GeV$}  
\begin{tabular}{|c|c|c|c|c|c|c|c|c|c|c|c|c|}
\hline mesons & \multicolumn{2}{c|}{ $\alpha_{s}=0.20$} & \multicolumn{2}{c|}{$\alpha_{s}=0.36$} & \multicolumn{2}{c|}{$\alpha_{s}=0.45$} & \multicolumn{2}{c|}{$\alpha_{s}=0.58$} & \multicolumn{2}{c|}{$\alpha_{s}=0.60$}  & \multicolumn{2}{c|}{$\alpha_{s}=0.64$}\\\hline 
 &\multicolumn{1}{l|}{$r_{1}$}& $r_{0}$ &\multicolumn{1}{l|}{$r_{1}$}&$r_{0}$&\multicolumn{1}{l|}{$r_{1}$}&$r_{0}$ &\multicolumn{1}{l|}{$r_{1}$}&$r_{0}$ &\multicolumn{1}{l|}{$r_{1}$}&$r_{0}$  &\multicolumn{1}{l|}{$r_{1}$}&$r_{0}$
 \\\hline
$D(c\bar{u}/c\bar{d})$&\multicolumn{1}{l|}{-}&-&\multicolumn{1}{l|}{-}&-&\multicolumn{1}{l|} {-}&- &\multicolumn{1}{l|} {6.990}&7.148 &\multicolumn{1}{l|} {6.757}&7.167& \multicolumn{1}{l|} {6.334}&7.204
\\\hline
$D(c\bar{s})$&\multicolumn{1}{l|}{-}&-&\multicolumn{1}{l|}{-}&-&\multicolumn{1}{l|}{-}&-&\multicolumn{1}{l|}{5.422}&6.151&\multicolumn{1}{l|}{5.241}&6.172&\multicolumn{1}{l|}{-}&-
\\\hline $B(\bar{b}u/\bar{b}d)$&\multicolumn{1}{l|}{-}&-&\multicolumn{1}{l|}{-}&{-}&\multicolumn{1}{l|}{-}&-&\multicolumn{1}{l|}{6.143}&6.151&\multicolumn{1}{l|}{5.938}&6.172&\multicolumn{1}{l|}{-}&- 
\\\hline $B_{s}(\bar{b}s)$&\multicolumn{1}{l|}{-}&-&\multicolumn{1}{l|}{-}&{-}&\multicolumn{1}{l|}{5.898
}&6.010&\multicolumn{1}{l|}{4.576}&6.151&\multicolumn{1}{l|}{-}&-&\multicolumn{1}{l|}{-}&- 
\\\hline $B_{c}(\bar{b}c)$&\multicolumn{1}{l|}{4.803}&5.719&\multicolumn{1}{l|}{2.668}&{5.908}&\multicolumn{1}{l|}{-
}&-&\multicolumn{1}{l|}{-}&-&\multicolumn{1}{l|}{-}&-&\multicolumn{1}{l|}{-}&-
\\\hline

\end{tabular}
\end{center}
\end{table}

\subsection{From the convergence Point of view} 
From the momentum transform , we see that for a lower cut-off value of $Q^{2}_{0}$, either one has to consider a very small value of $b$ or to increase the value of $\alpha_{s}$, which is obvious, since in both the cases coulombic part will be more dominant. However, reality condition of $\epsilon$ from equation $(10)$ demands that $\alpha_{s} \leq \frac{3}{4}$ and hence one can not go beyond $\alpha_{s}=0.75$ in this approach.\\

From the convergence point of view, the perturbative condition  demands \cite{NSB2009}
\begin{equation}
\frac{(4-\epsilon)(3-\epsilon)\mu b a^{3}_{0}}{2(1+a^{2}_{0}Q^{2})}<<C^{\prime}.
\end{equation} 

For a positive cut off ${Q_{0}}^{2}$, we can write
\begin{equation}
 \frac{(4-\epsilon)(3-\epsilon)\mu b a^{3}_{0}}{2(1+a^{2}_{0}{Q_{0}}^{2})}=C^{\prime}.
\end{equation}

The  values of $Q^{2}_{0}$ with $ b=0.183  GeV^{2} $for $ B $ and$ D $ mesons are shown in Table 2.

\begin{table}[h]
\begin{center}
\caption{Allowed range of $\alpha_{s}$ from the limit of $Q^{2}_{0}$ in the Model. }
\begin{tabular}{|c|c|c|c|c|c|c|}
\hline Mesons & $\alpha_{s}=0.20$  & $\alpha_{s}=0.36$ &$\alpha_{s}=0.45$&$\alpha_{s}=0.58$&$\alpha_{s}=0.60$&$\alpha_{s}=0.64$ 
\\\hline $D(\mu=0.2774 GeV)$ & $-$ & $-$ & $-$ & $0.01409$ &$0.01037$& $0.00204$
\\\hline $D_{s}(\mu=0.3576GeV)$ & $-$ &$-$ & $-$ & $0.00939$ & $0.00322$ & $-$
\\\hline  $B(\mu=0.3157GeV)$ & $-$ &$-$ & $-$ &$0.0125$ & $0.00775$ &$-$
\\\hline $B_{s}(\mu=0.4238GeV)$ & $-$ & $-$ & $0.04480$ &$0.00098$ &$-$ & $-$
\\\hline $B_{c}(\mu=1.171GeV)$ & $0.26378$ & $0.092543$ & $-$ &$-$ &$-$& $-$
\\\hline 
\end{tabular} 
\end{center}
\end{table}

Thus to incorporate lower value of $Q^{2}$ ($Q^{2}\leq\Lambda_{QCD}^{2}$), with linear part as perturbation, one expects a bound of $\alpha_{s} \leq 0.64$. 
%The variation of reduced mass $\mu$ and the cut off $Q^{2}_{0}$ for different  strong coupling constant $\alpha_{s}$ is shown in fig.1.

%\begin{figure}[h]
%\begin{center}
%\includegraphics[scale=0.8]{k.eps}  
%\caption {Upper limit of $\alpha_{s}$ for the different reduced masses of mesons $\mu$ and cut-off $Q_{0}^{2}$.}
%\end{center}
%\end{figure}

\subsection{Constraints on $\alpha_{s}$}
In the analysis, we further see that the value of $\alpha_{s}$ as well as  the model parameter `$c$'  also play a crucial role in choosing the parent and perturbative terms. From the above two constraints we make an individual range of $\alpha_{s}$ for different heavy light mesons and tabulate in Table 3. The range of `$\alpha_{s}$' and `$c$' within the two constraints are further specified in Figure 1.

\begin{table}[h]
\begin{center}
\caption{ Allowed range of {$\alpha_{s}$} and c for different mesons under the constraints. }
\begin{tabular}{|c|c|c|}
\hline Mesons & $\alpha_{s}$  & c  \\ 
\hline $D(\mu=0.2774 GeV)$ & $0.570-0.640$ &$\leq -1.2$\\ 
\hline $D_{s}(\mu=0.3576GeV)$ & $0.575-0.610$ &$-0.860$ to $-0.785$\\ 
\hline  $B(\mu=0.3157GeV)$ & $0.580-0.629$ &$-0.998$ to $-0.887$\\ 
\hline $B_{s}(\mu=0.4238GeV)$ & $0.450-0.582$ & $-0.997$ to $-0.663$\\ 
\hline$B_{c}(\mu=1.171GeV)$ & $0.200 - 0.409$ & $-0.994$ to $-0.197$\\
\hline 
\end{tabular} 
\end{center}
\end{table}

\begin{figure}[h]
\begin{center}
\includegraphics[scale=0.8]{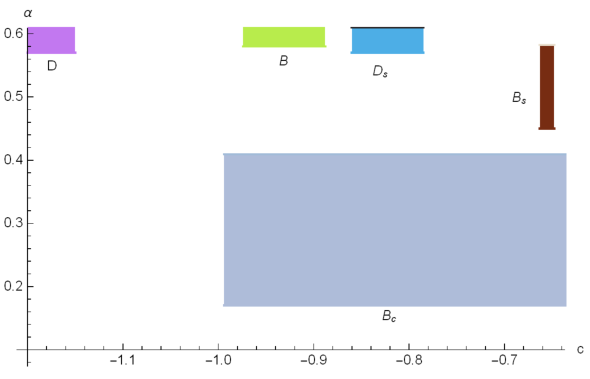}  
\caption {Range of $\alpha_{s}$ and $c$ for $D,D_{s}, B,B_{S}$ and $B_{c}$.}
\end{center}
\end{figure}.

\newpage

\section{Conclusions}  
In this work, we mainly focus in finding the analytical conditions to treat the linear part of the Cornell potential as perturbation. In the analysis we consider two constraints and evaluate a paramaterisation space for $\alpha_{s}$ and $c$ the range of $\alpha_{s}$ is found to be $0.20 \leq \alpha_{s}\leq 0.64$ with $-1.2 \leq c \leq -0.66$. We further note that the positive value of $c$ as is used in \cite{99} is excluded in the model and other values of $\alpha_s$ and `$c$' as is used in different phenomenolgical works \cite{faustov, 99, mao, NSB, CPL, NSB2009} are found to be valid within the parameterisation space provided in Figure 1. Further it is to be noted that the allowed space for $B_c$ meson is too large in comparison to other heavy-light mesons $D, D_s, B, B_s$ which may be due to the constituents of two heavy quarks and needs a detailed study to be carried out.  \\

However with linear part as perturbation, if the value of $\alpha_{s}$ in the above range is taken to be granted, then with the same potential another possibility of considering the coulombic part as perturbation also arises for a value of $\alpha_{s}\leq 0.20$.

\end{document}